\documentclass[a4paper,12pt]{article}
\usepackage{Stylefile}

\onehalfspace

\title
 {
\vspace*{2.0cm}
\LARGE{\bf Emissivity: A Program for Atomic Emissivity Calculations} \vspace*{5.0cm} \\
\author{Taha Sochi\footnote{University College London - Department of Physics \& Astronomy - Gower Street - London. Email:
t.sochi@ucl.ac.uk.} \vspace*{6.0cm}}
}

\makeindex

\begin{document}

\maketitle %
\pagenumbering{arabic}

\newpage
\phantomsection \addcontentsline{toc}{section}{Contents} %
\tableofcontents

\newpage
\phantomsection \addcontentsline{toc}{section}{Nomenclature}


\doublespace

{\setlength{\parskip}{6pt plus 2pt minus 1pt}

\pagestyle{headings} %
\addtolength{\headheight}{+1.6pt}
\lhead[{Chapter \thechapter \thepage}]%
      {{\bfseries\rightmark}}
\rhead[{\bfseries\leftmark}]%
     {{\bfseries\thepage}} 
\headsep = 1.0cm               

\newpage
\section{Abstract} \label{Abstract}

In this article we report the release of a new program for calculating the emissivity of atomic
transitions. The program, which can be obtained with its documentation from our website
\url{www.scienceware.net}, passed various rigorous tests and was used by the author to generate
theoretical data and analyze observational data. It is particularly useful for investigating atomic
transition lines in astronomical context as the program is capable of generating a huge amount of
theoretical data and comparing it to observational list of lines. A number of atomic transition
algorithms and analytical techniques are implemented within the program and can be very useful in
various situations. The program can be described as fast and efficient. Moreover, it requires
modest computational resources.

\newpage
\section{Program Summary} \label{Program Summary}

Title of the program: Emissivity \\
Type of program: command line \\
Programming language: C++ \\
Number of lines: 3110 \\
Computer: PC running Linux or Windows \\
Installation: desktop \\
Memory required to execute: case dependent \\
Speed of execution: case dependent ($\sim$ few seconds) \\
Has code been vectorized?: Yes \\
Has code been parallelized?: No \\
Compilers tested: g++, Dev-C++ \\
Number of warnings: 0 \\
Programming methodology: procedural with object oriented \\

\newpage
\section{Theoretical Background} \label{TheBac}
In a thermodynamic equilibrium situation an excited atomic state is populated by recombination and
radiative cascade from higher states, and depopulated by autoionization and radiative decay to
lower states. Many recombination lines arise from radiative decay and subsequent cascade of
strongly autoionizing resonance states near the ionization limit. These lines are dominated by low
temperature \dierec. The populations of such resonance states are determined by the balance between
autoionization and radiative decay. When autoionization dominates, the populations are then given
by the \Saha\ equation for thermodynamic equilibrium
\begin{equation}\label{SahaPop}
    \ND_{_{\BSRI}} = \NDe \ND_{_{\IECN}} \frac{\SW_{_{\BSRI}}}{2\SW_{_{\IECN}}}
    \left( \frac{\h^{2}}{2 \pi \me \BOLTZ \ElecT}  \right)^{3/2}
    e^{^{-\ED_{t} / \BOLTZ \ElecT}}
\end{equation}
where $\ED_{t}$ is the energy of the recombined electron in the $\BSRI$ state relative to the
ionization threshold, and the other symbols have their usual meaning as given in Nomenclature \S\
\ref{Nomenclature}.

The \Saha\ equation, which describes the ratio of different stages of ionization, is based on the
assumption of Local Thermodynamic Equilibrium (LTE) in a gas where collision dominates other
physical processes. Consequently, the local velocity and energy distributions of particles are
given by the Maxwell and Boltzmann distributions respectively and a temperature can be defined
locally. The \Saha\ equation is therefore strictly applicable only if elastic collisions are
responsible for establishing the energetic distribution of particles. In many practical cases,
however, atomic processes such as radiative transitions or dynamic effects are more important than
elastic collisions and the assumption of LTE is not justified within the whole energy range. In
these cases explicit detailed equilibrium calculations are required to determine the velocity and
energy distributions of particles over the various energy levels \cite{BenoyMS1993}.

To measure the departure of the state from thermodynamic equilibrium a \depcoe, $\DCu$, is defined
as the ratio of autoionization probability to the sum of radiative and autoionization
probabilities. The value of this coefficient is between 0 for radiative domination and 1 for
thermodynamic equilibrium. A large value of $\DCu$ ($\simeq 1$) is then required to justify the
assumption of a thermodynamic equilibrium and apply the relevant physics.

In thermodynamic equilibrium (TE) the rate of radiationless capture equals the rate of
autoionization, giving \cite{SeatonS1976}
\begin{equation}\label{depCoeBalTE}
    \NDe \NDi \RCc = \SPu \aTPu
\end{equation}
where $\NDe$ and $\NDi$ are the number density of electrons and ions respectively, $\RCc$ is the
\reccoe\ for the capture process, $\SPu$ is the \Saha\ population of the doubly-excited state and
$\aTPu$ is the autoionization transition probability of that state. In non-TE situation, the
balance is given by
\begin{equation}\label{depCoeBalnTE}
    \NDe \NDi \RCc = \NDu (\aTPu + \rTPu)
\end{equation}
where $\NDu$ is the non-TE population of the doubly-excited state and $\rTPu$ is the radiative
transition probability of that state. The \depcoe\ $\DCu$ is a measure of the departure from TE,
and hence is the ratio of the non-TE population to the \Saha\ population. Comparing Equations
(\ref{depCoeBalTE}) and (\ref{depCoeBalnTE}) gives
\begin{equation}\label{depCoe}
    \DCu = \frac{\NDu}{\SPu} = \frac{\aTPu}{\aTPu + \rTPu}
\end{equation}

\newpage
\section{Long Write-up} \label{LongWriteup}

\EmiCod\ code is a command line program that was developed to implement the atomic emissivity
model. Its main functionality is to calculate the emissivity of the transition lines from
electronic recombination and cascade decay all the way down to the ground or a metastable state,
that is all free-free, bound-free and bound-bound transitions. \EmiCod\ is written in C++ computer
language and mixes procedural with object oriented programming. It consists of about 3000 lines of
code. The program was compiled successfully with no errors or warnings using Dev-C++ compiler on
Windows, and g++ compiler on Cygwin and several Linux distributions such as Red Hat Enterprise.
Sample results produced by these three versions were compared and found to be identical. Elaborate
internal checks were carried out in all stages of writing and debugging the program and the output
was verified. Thorough independent checks on sample emissivity data produced by \EmiCod\ were
performed and found to be consistent.

The program reads from plain text input files and writes the results to a main plain text output
file. Other secondary output files can also be produced for particular purposes when required. The
necessary input files are
\begin{itemize}

\item The main file to pass the parameters and inform the program of the required calculations. The
main parameters in this file will be outlined in \S\ \ref{Input}.

\item A file for passing the free-states data and \oscstr s ($\OS$-values) for free-free and
bound-free transitions with photon energies. The free-states data include an index identifying the
resonance, its energy position, width, configuration, term, 2$J$, parity, and a flag marking the
energy position data as experimental or theoretical.

\item A file containing the bound-states data. These data include an index identifying the state, its
energy, \effquanum, configuration, term, and a flag marking the energy data as experimental or
theoretical.

\item A file containing the \oscstr s ($\OS$-values) for the bound-bound transitions.

\end{itemize}

The last file is imported from the \Rm-matrix code \cite{BerringtonEN1995, RmaxWeb} output while
the rest are user made. Two other input data files are also required if comparison and analysis of
observational data are needed. One of these is a data file that contains astronomical observations
while the other includes mapping information of observational lines to their theoretical
counterparts. The astronomical data file contains data such as observed wavelength, observed flux
before and after correcting for reddening and dust extinction, ionic identification, laboratory
wavelength, multiplet number, lower and upper spectral terms of transition, and statistical weights
of lower and upper levels. The mapping file contains the indices of the observational lines and
their theoretical counterparts.

In this section we summarize the theoretical background for emissivity calculations as implemented
in the \EmiCod\ code

\begin{itemize}

\item The program starts by obtaining the radiative transition probability $\rTPul$ for all
free-free transitions as given by
\begin{equation}\label{rTPul}
    \rTPul = \frac{\FSC^{3} \Ep^{2} \SWl \OSul}{2 \SWu \ATU}
\end{equation}
where the photon energy $\Ep$ is in \Ryd, and the other symbols are defined in Nomenclature \S\
\ref{Nomenclature}. This is followed by obtaining the radiative transition probability $\rTPul$ for
all free-bound transitions, as in the case of free-free transitions.


\item The total radiative transition probability $\rTP_{u}$ for all resonances is then found. This is the
probability of radiative decay from an upper resonance state to all accessible lower resonances and
bound states. This probability is found by summing up the individual probabilities $\rTPul$ over
all lower free and bound states $l$ for which a transition is possible according to the \eledip\
rules; that is
\begin{equation}\label{rTP}
    \rTPu = \sum_{l} \rTPul
\end{equation}


\item The \depcoe, $\DCu$, for all resonances is then obtained
\begin{equation}\label{departureCoef}
    \DCu = \frac{\aTPu}{\aTPu + \rTPu}
\end{equation}
where $\aTPu$ and $\rTPu$ are the autoionization and radiative transition probabilities of state
$u$, and $\aTP$ is given by
\begin{equation}\label{aTP}
    \aTP = \frac{\RW}{\Dirac}
\end{equation}


\item The next step is to calculate the population of resonances by summing up two
components: the \Saha\ capture, and the radiative decay from all upper levels. In thermodynamic
equilibrium (TE) the rate of radiationless capture equals the rate of autoionization, giving
\begin{equation}\label{depCoeBalTERes}
    \NDe \NDi \RCc = \SPl \aTPl
\end{equation}
where $\NDe$ and $\NDi$ are the number density of electrons and ions respectively, $\RCc$ is the
\reccoe\ for the capture process, $\SPl$ is the \Saha\ population of the doubly-excited state and
$\aTPl$ is the autoionization transition probability of that state. In non-TE situation, the
population and depopulation of the autoionizing state due to radiative decay from upper states and
to lower states respectively should be included, and hence the balance is given by
\begin{equation}\label{depCoeBalnTERes}
    \NDe \NDi \RCc + \sum_{u} \NDu \rTPul = \NDl (\aTPl + \rTPl)
\end{equation}
where $\NDl$ is the non-TE population of the doubly-excited state, $\rTPl$ is the radiative
transition probability of that state, $\rTPul$ is the radiative transition probability from an
upper state $u$ to the autoionzing state $l$, and the sum is over all upper states that can decay
to the autoionzing state. Combining (\ref{depCoeBalTERes}) and (\ref{depCoeBalnTERes}) yields
\begin{equation}\label{depCoeBalnTERes2}
    \SPl \aTPl + \sum_{u} \NDu \rTPul = \NDl (\aTPl + \rTPl)
\end{equation}
On manipulating (\ref{depCoeBalnTERes2}) the following relation can be obtained
\begin{eqnarray}\label{resonancePop2}
  \nonumber
  \NDl &=& \SPl \left( \frac{\aTPl}{\aTPl + \rTPl} \right) + \sum_{u} \frac{\NDu \rTPul}{\rTPl + \aTPl} \\
       &=& \SPl \DCl + \sum_{u} \frac{\NDu \rTPul}{\rTPl + \aTPl}
\end{eqnarray}
where $\DCl$ is the \depcoe\ of the autoionizing state. This last relation is used in calculating
the population.


\item The next step is to calculate $\rTPul$ for the bound-bound transitions. In these calculations the
$\OS$-values can be in length form or velocity form, though the length values are usually more
reliable, and hence \EmiCod\ reads these values from the $\OS$-values file produced by the
\Rm-matrix code. This is followed by finding $\rTPu$ for the bound states by summing up $\rTPul$
over all lower bound states $l$, as given earlier by (\ref{rTP}) for the case of resonances.


\item The population of the bound states is then obtained
\begin{equation}\label{boundPop}
    \NDl = \sum_{u} \frac{\NDu \rTPul}{\rTPl}
\end{equation}
where $u$ includes all upper free and bound states.


\item Finally, all possible free-free, free-bound and bound-bound transitions are found. The emissivity
of all recombination lines that arise from a transition from an upper state $u$ to a lower state
$l$ is then computed using the relation
\begin{equation}\label{emissivity1}
    \ENLul = \NDu \rTPul \h \F
\end{equation}
where $\F$ is the frequency of the transition line.


\item Apart from the normal debugging and testing of the program components to verify that they do what
they are supposed to do, two physical tests are incorporated within the program to validate its
functionality and confirm that no serious errors have occurred in processing and producing the
data. These tests are the population-balance test and the metastable test. The first test relies on
the fact that the population of each state should equal the depopulation. This balance is given by
the relation
\begin{equation}\label{firstTest}
    \sum_{j>i} \ND_{j} \rTP_{ji} = \ND_{i} \sum_{k<i} \rTP_{ik}
\end{equation}


The second test is based on the fact that the total number of the electrons leaving the resonances
in radiative decay should equal the total number arriving at the metastable states. This balance is
given by the relation
\begin{equation}\label{secondTest}
    \sum_{\forall j} \ND_{j} \rTP_{j} = \sum_{\forall i, k>i} \ND_{k} \rTP_{ki}
\end{equation}
where $i$ is an index for metastable states and $j$ is an index for resonances.

\end{itemize}


\section{Input and Output} \label{InputOutput}

The detailed technical description of the program input and output files is given in the program
documentation which can be obtained from \url{www.scienceware.net} website. However, in this
section we give a general description of the nature of the input parameters and the expected output
data so that the user can decide if the program is relevant for the required purpose.

\subsection{Input} \label{Input}

Various physical and computational parameters are required for successful run. Some of these are
optional and are only required for obtaining particular data. The main input parameters are

\begin{itemize}

\item The temperature in K and the number densities of electrons and ions in m$^{-3}$. The program uses a
temperature vector to generate data for various temperatures. These values can be regularly or
irregularly spaced.

\item The residual charge of the ion, which is required for scaling some of the data obtained from the
\Rm-matrix code output files.

\item Boolean flag for reading and processing observational astronomical data obtained from an external
input data file. A mapping file is also required for this process. If the user chooses to read
astronomical data, the data lines will be inserted between the theoretical lines in the main output
file according to their observed wavelength.

\item Boolean flag to run the aforementioned physical tests.

\item Parameters for generating and writing normalization emissivity data for theoretical and
observational lines to the main output file alongside the original emissivity data. The
normalization can be internal using the emissivity of one of the transition lines produced by the
program. It can also be with respect to an outside set of emissivity values corresponding to the
input temperatures. Another possibility is to normalize with respect to an outside set of
emissivity data in the form of effective \reccoe s corresponding to the specified input
temperatures. In this case the wavelength of the line to be normalized to is required. The
effective \reccoe\ $\RCf(\WL)$ is defined such that the emissivity $\EMISS(\WL)$ of a transition
line with wavelength $\WL$ is
\begin{equation}\label{EmissRecCoeff}
    \EMISS(\WL) = \NDe \NDi \RCf(\WL) \frac{\h \C}{\WL} \verb|    |   ({\rm J}.{\rm m}^{-3}.{\rm s}^{-1})
\end{equation}
Another choice for normalization is to be with respect to an outside set of emissivity data in the
form of effective \reccoe s corresponding to a set of temperatures that may be different from those
used in the input. In this case, the \reccoe s corresponding to the input temperature values are
obtained by interpolation or extrapolation using a polynomial interpolation routine.

\item Parameters to control the production and writing of the effective \reccoe s, $\RCf$,
which are equivalent to the given emissivities. Vacuum wavelength of the transition lines will be
used due to the restriction on the air wavelength formula ($\WL \ge 2000$\AA).

\item Parameters to control the algorithm to minimize the sum of least square deviations of
emissivity between the observational lines and their theoretical counterparts over the input
temperature range. This sum is given by
\begin{equation}\label{sumLSD}
    \mathcal{S} = \sum_{\forall \, \rm{lines}} \left(\EMISSoN - \EMISStN \right)^{2}
\end{equation}
where $\EMISSoN$ and $\EMISStN$ are the normalized observational and theoretical emissivities
respectively. This algorithm also offers the possibility of computing the temperature confidence
interval and its relevant parameters. The individual square differences for each one of the least
squares and their percentage difference which is given by
\begin{equation}\label{percentDiff}
    \left( \frac{\EMISSoN - \EMISStN}{\EMISSoN} \right) \times 100
\end{equation}
can also be written to the main output file. The confidence interval for the goodness-of-fit index
$\GoF$ can also be found if required by providing the degrees of freedom $\degFree$ and the change
in the goodness-of-fit index $\ChGoF$.

\item Parameters to run and control an algorithm for finding the decay routes to a particular state,
bound or free, from all upper states. As the number of decay routes can be very large (millions and
even billions) especially for the low bound states, a parameter is used to control the maximum
number of detected routes. The results (number of decay routes and the routes themselves grouped in
complete and non-complete) are written to an output file other than the main one. The states of
each decay route are identified by their configuration, term and 2$J$. For resonances, the \Saha\
capture term and the radiative decay term are also given when relevant. These temperature-dependent
data are given for a single temperature, that is the first value in the input temperature vector.

\end{itemize}

\subsection{Output} \label{Output}
The program produces a main output file which contains the essential emissivity data. The file
starts with a number of text lines summarizing the input data used and the output results, followed
by a few commentary lines explaining the symbols and units. This is followed by a number of data
lines matching the number of transitions. The data for each transition includes an index
identifying the transition, status (FF, FB or BB transition), the experimental state of the energy
data of the upper and lower levels, the attributes of the upper and lower levels (configuration,
term, 2$J$ and parity), wavelength in vacuum, wavelength in air for $\WL \geq 2000$\AA, the
emissivities corresponding to the given temperatures, the normalized emissivities and the effective
\reccoe s corresponding to the given emissivities. Writing the normalized emissivities and the
effective \reccoe s is optional and can be turned off. As mentioned earlier, if the user chooses to
read the astronomical data, the observational lines will be inserted in between the theoretical
lines according to their lab wavelength.

There are other subsidiary output files. One of these is a file containing the results of the
minimization algorithm. This file contains information on $\degFree$, $\ChGoF$, $\CISP$,
temperature at minimum $\GoF$, and the temperature limits for the confidence interval. This is
followed by the temperature array with the corresponding least squares residuals, the $\CISP \GoF$
and the $\GoF$ arrays as functions of temperature. Another output file is one containing mapping
data used for testing purposes. A third output file is a decay routes file which was outlined
previously.

\newpage
\section{Acknowledgement} \label{Acknowledgements}
The author would like to acknowledge the essential contribution of Prof. Peter Storey in all stages
of writing, testing and debugging the program and providing the theoretical framework. Without his
help and advice the development of the program would have been an impossible mission.

\newpage
\renewcommand{\refname}{}

\onehalfspace

\newpage
\section{Nomenclature and Notation} \label{Nomenclature}

\begin{supertabular}{ll}

$\FSC$          &       \finstr\ constant ($= \ec^{2}/(\Dirac \C \, 4\pi \PFS) \simeq 7.2973525376 \times 10^{-3}$) \\
$\aTP$          &       autoionization transition probability (s$^{-1}$) \\
$\rTP$          &       radiative transition probability (s$^{-1}$) \\
$\rTPul$        &       radiative transition probability from upper state $u$ to lower state $l$ (s$^{-1}$) \\
$\ChGoF$        &       change in goodness-of-fit index \\
$\RW$           &       width of resonance (J) \\
$\ED$           &       energy difference (J) \\
$\PFS$          &       permittivity of vacuum ($\simeq 8.854187817 \times 10^{-12}$F.m$^{-1}$) \\
$\EMISS$        &       emissivity of transition line (J.s$^{-1}$.m$^{-3}$) \\
$\ENLul$        &       emissivity of transition line from state $u$ to state $l$ (J.s$^{-1}$.m$^{-3}$) \\
$\EMISSoN$      &       normalized observational emissivity \\
$\EMISStN$      &       normalized theoretical emissivity \\
$\degFree$      &       number of degrees of freedom \\
$\WL$           &       wavelength (m) \\
$\F$            &       frequency (s$^{-1}$) \\
$\RCc$          &       \reccoe\ for capture process (m$^{3}$.s$^{-1}$) \\
$\RCf$          &       effective \reccoe\ (m$^{3}$.s$^{-1}$) \\
$\ATU$          &       atomic time unit ($= \Dirac / \Eh \simeq 2.418884326505 \times 10^{-17}$s) \\
$\GoF$          &       goodness-of-fit index \\
%
 \vspace{-0.3cm} \\
%
%
$\CISP$         &       scaling parameter in $\GoF$ confidence interval procedure \\
{\AA}           &       angstrom \\
$\DC$           &       \depcoe\ \\
$\DCu$          &       \depcoe\ of upper state\\
$\C$            &       speed of light in vacuum (299792458 m.s$^{-1}$) \\
$\ec$           &       elementary charge ($\simeq 1.602176487 \times 10^{-19}$C) \\
$\ele$          &       electron \\
$\E$            &       energy (J) \\
$\Eh$           &       \Hartree\ energy ($\simeq 4.35974394 \times 10^{-18}$J) \\
$\Ep$           &       photon energy (J) \\
$\OS$           &       \oscstr \\
$\OSul$         &       \oscstr\ of transition between upper state $u$ and lower state $l$ \\
$\SW$           &       statistical weight in coupling schemes ($=2J+1$ for IC) \\
$\h$            &       Planck's constant ($\simeq \Sci {6.6260693}{-34}$ J.s) \\
$\Dirac$        &       reduced Planck's constant ($= h/2\pi \simeq \Sci {1.0545717}{-34}$ J.s) \\
$J$             &       total angular momentum quantum number \\
\BOLTZ          &       \Boltzmann's constant ($\simeq \Sci {1.3806505}{-23}$ J.K$^{-1}$) \\
$L$             &       total orbital angular momentum quantum number \\
$\m$            &       mass (kg) \\
$\me$           &       mass of electron ($\simeq 9.10938215 \times 10^{-31}$kg) \\
$\ND$           &       number density (m$^{-3}$) \\
$\NDe$          &       number density of electrons (m$^{-3}$) \\
$\NDi$          &       number density of ions (m$^{-3}$) \\
$\SP$           &       \Saha\ population (m$^{-3}$) \\
\Rm             &       resonance matrix in \Rm-matrix theory \\
$r$             &       radius (m) \\
$S$             &       total spin angular momentum quantum number \\
$\ti$           &       time (s) \\
$\T$            &       temperature (K) \\
$\ElecT$        &       electron temperature (K) \\
$\IECN$         &       ion of effective positive charge $n$ \\
$\BSRI$         &       bound state of recombined ion with effective positive charge $n-1$ \\
%
$\verb|      |$ & \\
\end{supertabular}

\VS

{\LARGE \bf \noindent Abbreviations} \vspace{0.3cm} \\
\begin{supertabular}{ll}
%

BB              &       Bound-Bound \\
C               &       \Coulomb \\
F               &       Faraday \\
FB              &       Free-Bound \\
FF              &       Free-Free \\
IC              &       \IntCou \\
J               &       Joule \\
K               &       Kelvin \\
kg              &       kilogram \\
LTE             &       Local Thermodynamic Equilibrium \\
(.)$_{l}$       &       lower \\
m               &       meter \\
\Ryd            &       \Rydberg \\
s               &       second \\
TE              &       Thermodynamic Equilibrium \\
(.)$_{u}$       &       upper \\
(.)$_{ul}$      &       upper to lower \\
$\forall$       &       for all \\
$\verb|      |$ & \\

\end{supertabular}

\vspace{0.5cm}

\noindent %
{\bf Note}: units, when relevant, are given in the SI system. Vectors are marked with boldface.
Some symbols may rely on the context for unambiguous identification. The physical constants are
obtained from the National Institute of Standards and Technology (\NIST) website \cite{NIST2007}.
It should be remarked that most subscripts (e.g. $l$ and $u$) are dummy variables, that is they are
subject in their interpretation to the context and are not specific to the particular cases they
are referring to.

\end{document}